\documentclass[osajnl2,twocolumn,showpacs,superscriptaddress]{revtex4}  
\pdfoutput=1
\usepackage[usenames,dvipsnames]{color}
\usepackage{amsmath,amssymb}
\usepackage[colorlinks=true, citecolor=blue, linkcolor=blue]{hyperref}

\makeatletter

\begin{document}

\title{Multi-color Cavity Metrology}

\author{Kiwamu Izumi}
\affiliation{Department of Astronomy, Graduate school of science, University of Tokyo, Bunkyo-ku,
 Hongo, Tokyo 113-0033, Japan}
\author{Koji Arai}
\affiliation{LIGO Laboratory, California Institute of Technology MS 100-36, Pasadena, CA 91125, USA}
\author{Bryan Barr}
\affiliation{SUPA, School of Physics \& Astronomy, University of Glasgow, Glasgow, G12 8QQ, Scotland}
\author{Joseph Betzwieser}
\affiliation{LIGO Livingston Observatory, P.O. Box 940, Livingston, Louisiana 70754-0940, USA}
\author{Aidan Brooks}
\affiliation{LIGO Laboratory, California Institute of Technology MS 100-36, Pasadena, CA 91125, USA}
\author{Katrin Dahl}
\affiliation{Max-Planck-Institut f\"{u}r Gravitationsphysik (Albert-Einstein-Institut) and Leibniz Universit\"{a}t Hannover, Callinstr. 38, 30167 Hannover, Germany}
\author{Suresh Doravari}
\affiliation{LIGO Laboratory, California Institute of Technology MS 100-36, Pasadena, CA 91125, USA}
\author{Jennifer C. Driggers}
\affiliation{LIGO Laboratory, California Institute of Technology MS 100-36, Pasadena, CA 91125, USA}
\author{W. Zach Korth}
\affiliation{LIGO Laboratory, California Institute of Technology MS 100-36, Pasadena, CA 91125, USA}
\author{Haixing Miao}
\affiliation{LIGO Laboratory, California Institute of Technology MS 100-36, Pasadena, CA 91125, USA}
\author{Jameson Rollins}
\email[Corresponding author: ]{jrollins@ligo.caltech.edu}
\affiliation{LIGO Laboratory, California Institute of Technology MS 100-36, Pasadena, CA 91125, USA}
\author{Stephen Vass}
\affiliation{LIGO Laboratory, California Institute of Technology MS 100-36, Pasadena, CA 91125, USA}
\author{David Yeaton-Massey}
\affiliation{LIGO Laboratory, California Institute of Technology MS 100-36, Pasadena, CA 91125, USA}
\author{Rana X. Adhikari}
\affiliation{LIGO Laboratory, California Institute of Technology MS 100-36, Pasadena, CA 91125, USA}

\begin{abstract}
  Long baseline laser interferometers used for gravitational wave
  detection have proven to be very complicated to control.  In order
  to have sufficient sensitivity to astrophysical gravitational waves,
  a set of multiple coupled optical cavities comprising the
  interferometer must be brought into resonance with the laser
  field. A set of multi-input, multi-output servos then lock these
  cavities into place via feedback control. This procedure, known as
  lock acquisition, has proven to be a vexing problem and has reduced
  greatly the reliability and duty factor of the past generation of
  laser interferometers. In this article, we describe a technique for
  bringing the interferometer from an uncontrolled state into
  resonance by using harmonically related external fields to provide a
  deterministic hierarchical control. This technique reduces the
  effect of the external seismic disturbances by four orders of
  magnitude and promises to greatly enhance the stability and
  reliability of the current generation of gravitational wave
  detector.  The possibility for using multi-color techniques to
  overcome current quantum and thermal noise limits is also discussed.
\end{abstract}


\maketitle

\section{Introduction}
Gravitational waves promise to reveal new information about the bulk
motions of massive compact objects in the universe.  In this decade
kilometer-scale interferometers, such as LIGO~\cite{LIGO:Science,
  PF:RPP2009}, Virgo~\cite{Virgo:2011},
GEO600~\cite{0264-9381-25-11-114043}, and
KAGRA~\cite{0264-9381-27-8-084004}, are expected to make the first
direct detection of gravitational waves in the 10-10k~Hz band.  The
Advanced LIGO (aLIGO) project~\cite{0264-9381-27-8-084006} is a
significant upgrade of the initial LIGO interferometers, including
more sophisticated vibration isolation, a factor of ten higher laser
power, larger test masses, and a more versatile optical readout, among
other improvements.  These improvements should lead to a factor of 10
sensitivity improvement across the entire detection band, resulting in
a factor of 1000 in increased probed volume of space.

The problem of moving an interferometer from its initial uncontrolled
state (where the suspended mirrors are swinging freely) to the final
operating state (where all cavity lengths are interferometrically
controlled) is referred to as ``lock acquisition''.  For a single
Fabry-Perot cavity or a simple Michelson interferometer, the problem
is relatively straightforward: typically the locking servo is engaged
and the mirrors are moved until the servo forces the interferometer
into the desired operating point.  With a more complicated
configuration utilizing multiple coupled cavities (e.g. aLIGO) there
are no reliable optical signals providing cavity length information
until all cavities are simultaneously resonant. This problem is
compounded by seismic-induced residual mirror motions which, even with
the advanced seismic isolation systems used in aLIGO, are expected to
be $\sim\!10^{-7}$~m below 1~Hz~\cite{strain:044501}.  Waiting for
full acquisition to happen by chance is an exercise in futility.

Over the years various techniques have been developed to address the
lock acquisition problem.  Algorithms have been developed that use
digital controls, clever sequencing of feedback loops and mixing of
interferometric signals to reduce the waiting time~\cite{Camp:95,
  Evans:02, Virgo:Lock}.  Unfortunately, these are still insufficient
for the needs of aLIGO.  Instead, aLIGO will be using an auxiliary arm
length stabilization system to robustly bring the long arm cavities to
a stable operating point at the cavity resonance, independent of the
rest of the interferometer.  This technique uses frequency-doubled
auxiliary lasers, phase locked to the main interferometer laser
source, to serve as an independent sensor for the cavity lengths.  The
primary motivation of the experiment described in this article is to
demonstrate that the arm cavity length can be independently controlled
by auxiliary locking to within a small fraction of the linewidth of
the arm cavity at the primary laser frequency.

Beyond the practical application of aLIGO arm length stabilization,
this type of ``multi-color metrology'' can allow us to make better
measurements of cavity properties and noise fluctuations.  This may
one day make it possible to sense inherent noise sources and reduce
their metrological effects.

Arm length stabilization with frequency-doubled auxiliary lasers has
been demonstrated by Mullavey et. al~\cite{Mullavey:12} in
shorter-baseline cavities.  Other methods for independent arm cavity
stabilization have included \textit{digital}
interferometry~\cite{Shaddock:07, Lay:07}, and \textit{suspension
  point} (or \textit{suspension platform})
interferometry~\cite{Aso2004, Numata:08}.

This article describes a prototype experiment of cavity length
stabilization using multiple laser wavelengths, and its implications
for future interferometers.  The experiment has been performed on a
40-meter-long suspended Fabry-Perot cavity.  Possible noise sources
and their contributions are discussed.

\section{Experimental Setup}
\label{sec:setup}

This experiment was conducted on the Caltech-LIGO 40m prototype
interferometer~\cite{Osamu:Lock, PhysRevD.74.022001, Rob:2008}.  This
prototype is used to develop interferometer topologies for future
gravitational wave detectors.  Currently the configuration is similar
to that of aLIGO (dual-recycled Michelson with 40m-long Fabry-Perot
arm cavities), and is being used to prototype aLIGO readout and
control schemes.  ``Dual-recycling'' refers to the use of both power
and signal recycling mirrors at the input and output ports
respectively of the Michelson interferometer, recycling light power
that would otherwise escape through those ports~\cite{Mee1988,
  MiEa1993}.  All of the main interferometer optics are suspended as
single-stage pendula with a length of 25~cm ($f_p = 1$~Hz).  The
entire interferometer is housed in an ultra-high vacuum envelope.  For
this experiment just a single, suspended, 40m-long Fabry-Perot arm
cavity of the full interferometer is used.  The rest of the
interferometer optics are misaligned so as to not affect the
measurements.

\begin{figure*}
  \centering
  \includegraphics[width=2\columnwidth]{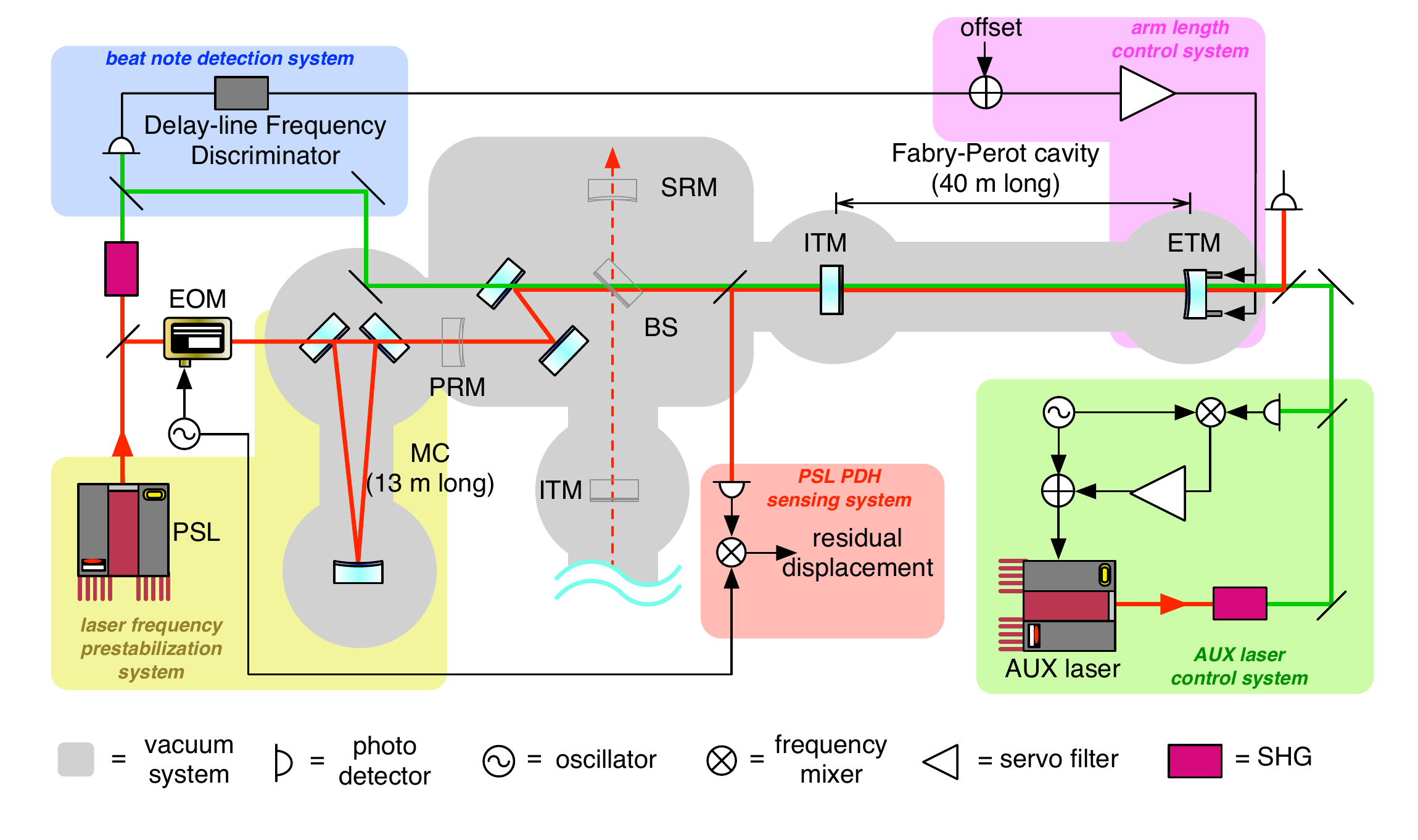}
  \caption{Experimental setup.  Red lines indicate the path of the
    1064~nm PSL beam, and green lines indicate the path of the 532~nm
    AUX beam.  The colored regions correspond to logical sections of
    the control and readout, described in more detail in
    Section~\ref{sec:model}. Optics in shadow are part of the larger
    interferometer not used in this experiment.}
  \label{fig:setup}
\end{figure*}

Figure~\ref{fig:setup} shows a schematic diagram of the experimental
setup.  A 1064~nm beam (red in diagram) from the main interferometer
pre-stabilized laser (PSL) is injected into the arm cavity through the
input mirror (ITM) from the interferometer vertex.  A second, 532~nm
beam (green in diagram) from an auxiliary (AUX) laser is injected
through the cavity end mirror (ETM).  Whereas the PSL beam circulates
through the full interferometer under normal operating conditions, the
AUX laser beam resonates only in the single arm cavity and is
extracted before interacting with the rest of the interferometer.

The cavity mirrors are dichroic and highly reflective at both
wavelengths.  For the 1064~nm beam, the cavity is overcoupled and
reflects most of the light back towards the interferometer vertex.
For the 532~nm beam the cavity has a much higher transmissivity and
some of the light is transmitted through the cavity at the ITM,
through the interferometer vertex and extracted from the vacuum.  The
mirror and cavity properties for both wavelengths are shown in
Table~\ref{tab:cavity}.

\begin{table}
\begin{center}
\begin{tabular}{cccc}
  \hline
  \hline
  \bf cavity property & \bf symbol & \bf 1064 nm & \bf 532 nm \\
  \hline
  ITM power transmissivity & $T_i$ & $0.0138$ & $0.0458$ \\
  ETM power transmissivity & $T_e$ & $1.37\!\times\!10^{-5}$ & $0.0109$ \\
  power trans. (resonance) & $T_c$ & $3.92\!\times\!10^{-3}$ & $0.616$ \\
  power trans. (anti-resonance) & $T_c^\dagger$ & $4.77\!\times\!10^{-8}$ & $1.29\!\times\!10^{-4}$ \\
  finesse & $\mathcal{F}$ & 450 & 109 \\
  cavity pole frequency & $f_c$ & 4.40 kHz & 18.3 kHz \\
  cavity length& $L$ & \multicolumn{2}{c}{37.8 m} \\
  free spectral range & $f_{\rm FSR}$ &  \multicolumn{2}{c}{3.97 MHz} \\
  \hline
  \hline
\end{tabular}
\caption{Cavity properties of the arm cavity, at 1064~nm as seen 
from the vertex, and 532~nm as seen from cavity end.}
\label{tab:cavity}
\end{center}
\end{table}

\subsection{PSL light source}
The 1064~nm light source is a 2~W Innolight NPRO.  The light is
spatially filtered by a $\sim\!20$~cm ring cavity with a $\sim\!2$~MHz
bandwidth that also provides passive filtering of the laser noise at
RF frequencies. The PSL frequency is locked to an in-vacuum,
suspended, critically-coupled, triangular mode cleaner (MC) cavity,
which conditions the beam by suppressing excess frequency noise and
rejecting higher-order spatial modes.  The laser is locked to the MC
via the standard Pound-Drever-Hall (PDH) method~\cite{DrEA1983} with a
$\sim\!130$~kHz bandwidth.  The power is adjusted to allow
approximately 25~mW of 1064~nm laser light to be incident on the
cavity under test.

\subsection{AUX light source and frequency doubling}
The AUX beam comes from a frequency-doubled 700~mW JDSU NPRO-126N.
The frequency doubling is achieved via second harmonic generation
(SHG) in a PPKTP crystal~\cite{Greenstein2004319}.  The conversion
efficiency is $\sim\!1\%/\mathrm{W}$, and with other input losses we
end up with 1.2~mW of 532~nm light incident on the ETM.

\subsection{Dichroic mirror coatings}
\begin{figure}
  \centering
  \includegraphics[width=\columnwidth]{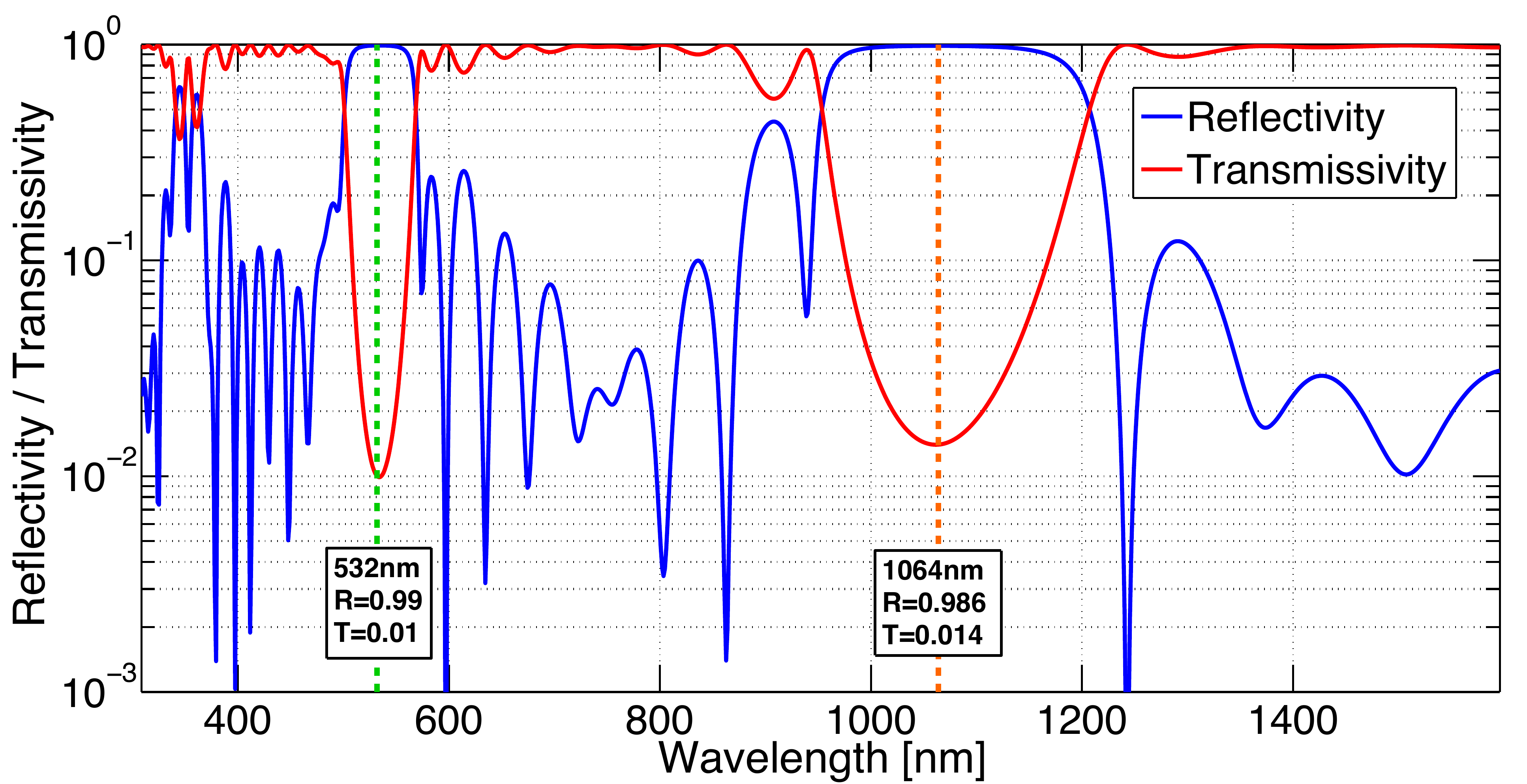}
  \caption{Spectral reflectivity of one of the dichroic cavity mirrors. }
  \label{fig:coat}
\end{figure}

The mirrors forming the Fabry-Perot arm cavities have custom coatings
to provide reflectivity at both 1064 and 532~nm.
Figure~\ref{fig:coat} shows the calculated coating reflectivity as a
function of wavelength for the ITM (the ETM shows a similar
profile). The layer structure is a particular aperiodic design chosen
to minimize the influence of various types of thermal noise on the
reflected phase of the laser field~\cite{Matt:TOnoise}.

\subsection{Sensing, acquisition and control}
\label{sec:control}
Initially, the AUX laser frequency is locked to the arm cavity length
via a standard PDH locking scheme.  The AUX laser is locked to the
cavity, rather than vice versa, because the laser frequency actuator
has much greater bandwidth than the cavity mirror displacement
actuators.  Phase modulation sidebands at 217~kHz are introduced on
the AUX beam by directly driving the laser frequency actuator.  This
modulation frequency is chosen to minimize the ratio of amplitude to
phase modulation.  The green light reflected from the ETM is used for
the PDH lock.

Once the AUX laser is locked to the cavity, the AUX beam transmits
through the ITM and is extracted from the vacuum system.  The
extracted AUX beam is used in a heterodyne measurement with a
frequency-doubled sample of the PSL beam.  The frequency of the beat
note between the AUX laser and the frequency-doubled PSL is measured
by the a delay-line frequency discriminator (DFD) (see
section~\ref{sec:dfd}).  The DFD has ``coarse'' and ``fine'' paths
which provide different dynamic ranges.  These outputs are the primary
error signal for the cavity control.  They are digitized and a control
signal is generated with a digital feedback control system.

\begin{figure*}
  \centering
  \includegraphics[width=2\columnwidth]{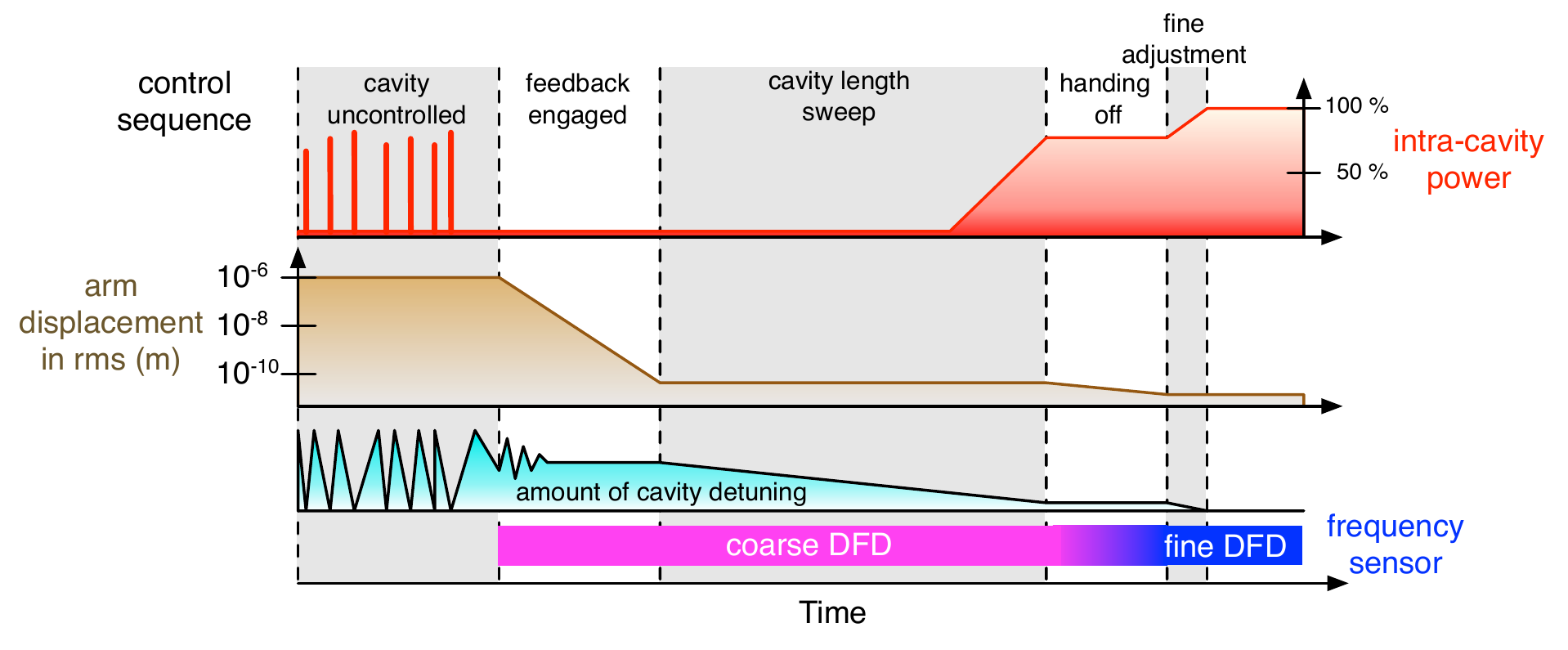}
  \caption{Sequence of the arm length control as a function of time.
    The intra-cavity power and detuning are for the 1064~nm PSL beam.}
  \label{fig:sequence}
\end{figure*}

Figure~\ref{fig:sequence} shows the control sequence and hand-off
between the coarse and fine discriminator paths.  Since seismic noise
acting on the length of the MC and arm cavities causes the beat note
to fluctuate by about 10~MHz, the large range coarse path is used to
engage feedback smoothly.  After length control is achieved an
artificial offset is introduced in the discriminator signal to sweep
the length of the arm cavity until the length meets the resonance
condition for the PSL beam.  This ability to tune the cavity length
directly is the key to the use of this technique as a lock acquisition
tool for Advanced LIGO.

In the end, control is passed to the fine discriminator by digitally
fading over between the coarse and fine signals.  The fine
discriminator has a higher signal-to-noise by a factor of the extra
delay.  At this stage the arm length can be tuned more precisely so
that the main laser fully resonates in the arm.  The final
steady-state control used during the measurement is described in
detail in section~\ref{sec:model}.  A PDH error signal derived from
the PSL beam reflected off of the ITM provides an out-of-loop measure
of the residual cavity displacement.

\subsection{Delay-line frequency discriminator}
\label{sec:dfd}
The delay-line frequency discriminator (DFD), which is used to measure
the frequency of the beat note between the PSL and AUX beams, works by
mixing an input RF signal with a delayed version of itself.  For a
given delay in the delay line, $D$, the mixer output voltage is a
periodic function of the input frequency, $f$.  In the small frequency
limit ($f \ll 1/D$) the output is directly proportional to the input
frequency:
\begin{equation}
  V \propto D f.
  \label{eq:dfd}
\end{equation}

\begin{figure}
  \centering
  \includegraphics[width=\columnwidth]{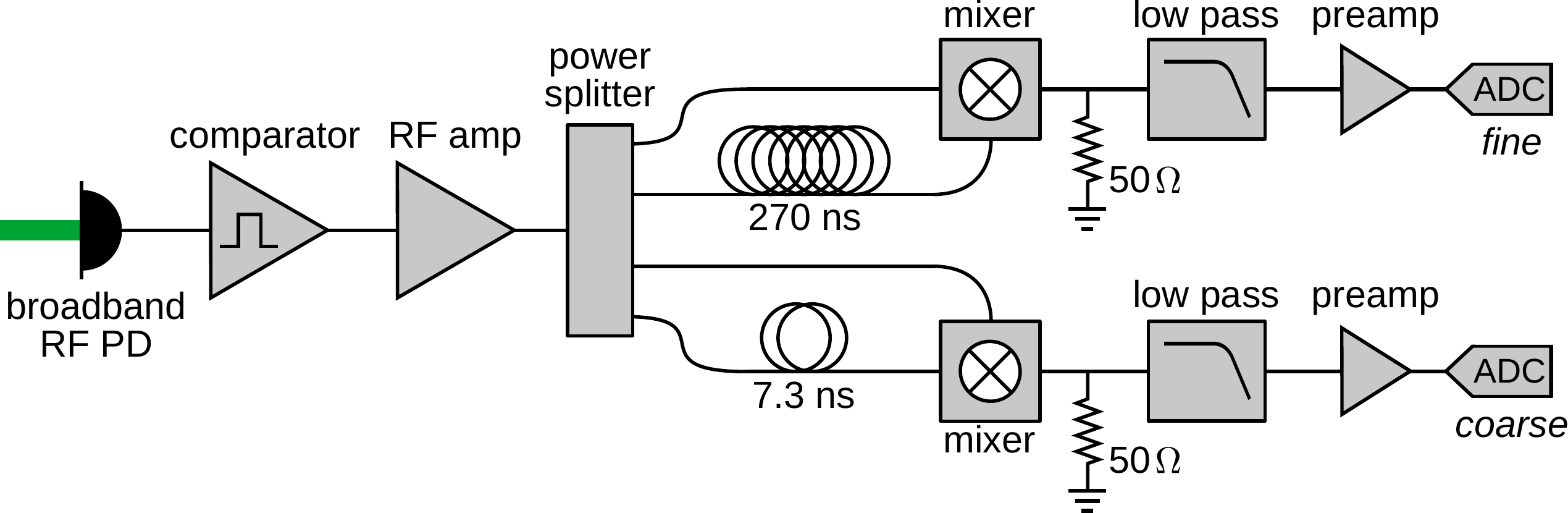}
  \caption{Delay-line frequency discriminator.  Components used:
    comparator: Analog Devices AD9696; RF amplifier: Mini-Circuits
    ZHL-1A; power-splitter: Mini-Circuits ZBSC-413+; delay line
    cables: RG-58 C/U; mixer: Mini-Circuits ZP-3+; low-pass filters:
    Mini-Circuits BLP-1.9; pre-amplifier: Stanford Research SR560.}
  \label{fig:dfd}
\end{figure}

Figure~\ref{fig:dfd} shows a schematic of the DFD circuit.  The signal
from a broadband RF photodetector first passes through a comparator
that turns the signal into a square wave.  This helps reduce noise
associated with small amplitude fluctuations of the input signal.
This signal is amplified and split into two discriminator paths: a
\textit{coarse} path with a delay of 7.3~ns and frequency range of
34~MHz, and a \textit{fine} path with a delay of 270~ns and a range of
3.6~MHz.  The mixer outputs, with signals given by
equation~\ref{eq:dfd}, are filtered and digitized and used as error
signals for the cavity length control servo.  The coarse path, which
has a larger bandwidth, is used during lock acquisition, whereas the
lower-noise fine path is used to achieve best performance in the
steady state.

In order to confine the cavity length to the within the line width of
the PSL a residual fluctuation level of 10~pm RMS must be achieved.
This means that the frequency noise of the fine path needs to be less
than 7.4~Hz$/\sqrt{\mathrm{Hz}}$ in the control bandwidth.  The noise
of the two paths are currently limited by the active readout
electronics at an estimated level of 2.0~Hz/$\sqrt{\mathrm{Hz}}$ and
0.1~Hz/$\sqrt{\mathrm{Hz}}$ for the coarse and fine paths
respectively.  They therefore reasonably meet the required frequency
stability.

A delay-line design was used, rather than a phase-locked loop (PLL)
design~\cite{schilt:123116} because the frequency range of DFDs are
relatively easy to tune and can be adjusted to give a large frequency
range (e.g. the coarse channel).  DFDs also don't require any active
feedback loops, which complicate PLLs.  Alternatively, a combination
of a large range DFD and a smaller range PLL could be a possible
solution depending on the required frequency range and noise.

\section{Control Model}
\label{sec:model}

\begin{figure*}
  \centering
  \includegraphics[width=2\columnwidth]{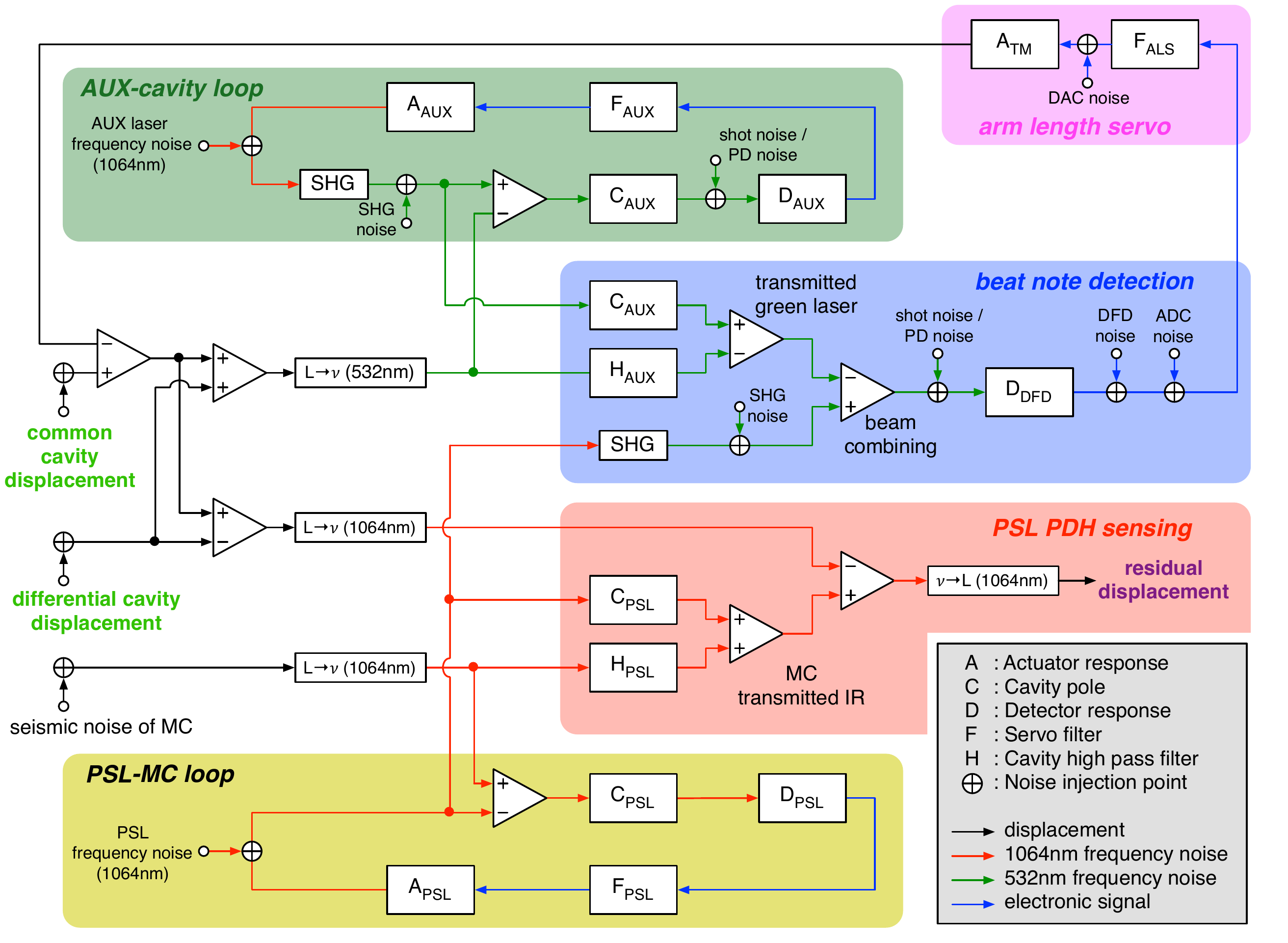}
  \caption{Block diagram of the model of control scheme and noise
    sources for the experiment.  The colored blocks correspond to the
    colored blocks in the setup diagram in Figure~\ref{fig:setup}.
    The various individual control elements are described in the text
    and in appendix~\ref{sec:zpk}.}
    \label{fig:model}
\end{figure*}

In this section we present a model of the control system used in the
experiment.  The control system, shown in Figure~\ref{fig:model}, can
be broken into five parts, each described in the sections below.  The
model includes injection points for various noise sources that might
affect overall performance, discussed in detail in
section~\ref{sec:noise}.

\subsection{AUX-cavity loop}
The first logical control loop is the PDH lock of the
frequency-doubled AUX laser to cavity (green block in
Figures~\ref{fig:setup} and~\ref{fig:model}).  This loop suppresses
the frequency noise of the AUX laser and allows its frequency to
follow the motion of the cavity length.  The control bandwidth of this
loop is 30~kHz, limited by the laser cavity PZT frequency actuator.

Since the AUX laser is locked to the cavity, information about the
length fluctuation of the cavity is encoded in the frequency of the
AUX laser light transmitted through the cavity ITM.  The relationship
between cavity length and laser frequency is given by:
\begin{equation}
  \frac{ d L}{L} = \frac{d \nu}{\nu},
  \label{eq:f2l}
\end{equation}
where $L$ is the cavity length and $\nu$ is the frequency of the laser
resonating in the cavity.  The cavity frequency response is a function
of the cavity finesse and can be approximated as a single-pole low
pass filter:
\begin{equation}
  C_{\mathrm{AUX}}(f) = \frac{1}{1 + \mathrm{i} ( f/f_c)},
  \label{eq:lpf}
\end{equation}
where $f$ is the frequency of the signal and $f_c$ is the cavity pole
frequency, which for the 532~nm AUX beam is 18~kHz.

The photo detection and mixing process that produces the PDH error
signal has an overall flat V/Hz conversion factor given by
$D_{\mathrm{AUX}}$.  The servo filter, which is tuned to provide
stable and robust locking, has a frequency response of
$F_{\mathrm{AUX}}$. Finally, the laser PZT frequency actuator has a
response of $A_{\mathrm{AUX}}$.

The noise sources associated with this loop are the AUX laser
frequency noise, second-harmonic generation noise, shot noise at the
detector, and electronics noise of the readout electronics.  These
noise sources will be discussed in detail in section~\ref{sec:noise}.

\subsection{PSL-MC loop}
\label{sec:pslmc}
The PSL-MC loop (yellow in Figures~\ref{fig:setup}
and~\ref{fig:model}) describes the lock of the PSL frequency to the
mode cleaner length.  The control bandwidth of this loop is 130~kHz.
As with the AUX laser cavity loop, the MC cavity has a single-pole
frequency response given by $C_{\mathrm{PSL}}$.  $D_{\mathrm{PSL}}$ is
the response of the PDH sensing, $F_{\mathrm{PSL}}$ is the response of
the servo filter, and $A_{\mathrm{PSL}}$ is the response of the PSL
compound frequency actuator, which includes the laser crystal
temperature actuator, laser cavity PZT, and an EOM phase modulator.

The PSL light transmitted through the MC and incident on the arm
cavity under test is the reference for the performance of the cavity
stabilization system.  However, finite gain in the PSL-MC loop can
potentially lead to PSL frequency noise coupling into the arm
stabilization loop.  Despite this, the model shows that the
suppression ratio from PSL frequency noise to residual displacement is
more than $10^6$ at 100~Hz, so we can safely neglect the PSL frequency
noise.

\subsection{Beat note detection}
The beat note detection block (blue in Figures~\ref{fig:setup} and
~\ref{fig:model}) measures the frequency difference between the AUX
and frequency-doubled PSL beams.

The transmitted AUX light is a combination of the suppressed AUX laser
frequency noise and any external displacement noises in the cavity
that modulate the optical phase of the laser resonating in the cavity.
These external disturbances result in phase noise that is converted to
frequency noise by multiplying by the complex frequency,
$\mathrm{i}f$.  This signal is then low-pass filtered by the cavity
pole, $C_{\mathrm{AUX}}$, with a resultant transfer function of:
\begin{equation}
  H_{\mathrm{AUX}}(f) = \frac{\mathrm{i} (f/f_c)}{1 + \mathrm{i} (f/f_c)},
\end{equation}
where again $f_c = 18$~kHz.  $D_{\mathrm{DFD}}$ represents the flat
Hz$\rightarrow$V conversion of the full beat detection process,
including the gain of the RF photo detector and the delay-line
frequency discriminator.

Noise in the frequency doubling process of the PSL should be at a
similar level to that in the AUX-cavity loop.  Laser shot and
photo-detection noises are also similar to those in AUX-cavity loop.
The electronics noise in this case is from the frequency discriminator
(DFD).  Finally, there is also additional noise from the
analog-to-digital conversion (ADC) process.

\subsection{Arm length servo}
The output of the beat note detection process is the error signal for
the arm cavity length control servo loop (pink in
Figures~\ref{fig:setup} and~\ref{fig:model}).  The digital error
signal is sent through the servo filter, $F_{\mathrm{ALS}}$, which
includes a 470~$\mu$s processing delay.  The resultant digital control
signal is converted back to an analog voltage via a digital-to-analog
converter (DAC), and the output analog control signal is used to
actuate on the end test mass via electro-magnetic actuators
($A_{\mathrm{TM}}$).  The overall open loop gain of this loop is
roughly
\begin{equation}
  G_{\mathrm{ALS}} \simeq D_{\mathrm{DFD}}\,F_{\mathrm{ALS}}\,A_{\mathrm{TM}},
  \label{eq:Gals}
\end{equation}
since the effect of the AUX-cavity loop and $H_{\mathrm{AUX}}$ in
parallel is an overall flat frequency response that does not affect
the overall open-loop gain.

When the loop is closed, fluctuations in the frequency of the beat
note are suppressed by acting on the cavity length.  Any external
disturbances that produce frequency shifts common to both the AUX and
PSL beams will then be suppressed by the closed loop suppression
factor, $1/(1+G_{\mathrm{ALS}})$.  The transfer function between
disturbances common to both wavelengths and residual displacement can
be seen in Figure~\ref{fig:bode} in appendix~\ref{sec:zpk}.

\subsection{PSL PDH sensing}
\label{sec:pslpdh}
This block represents the direct out-of-loop measurement of the
residual displacement noise of the cavity relative to the length of
the MC.  As opposed to the external disturbances common to the AUX and
PSL beams that are suppressed by the arm length servo, any external
disturbances sensed \textit{differentially} between the AUX and PSL
beams will transmit directly to this sensor and contribute to any
residual displacement noise.  The transfer function between
differential external disturbances and residual displacement can be
seen in Figure~\ref{fig:bode} in appendix~\ref{sec:zpk}.

There are noise sources here related to readout, such as ADC noise,
and shot noise and dark noise in the MC transmitted photo detection
process, but they are found to be insignificant relative to other
noise sources and are therefore omitted.

\section{Performance and Noise Analysis}
\label{sec:noise}
\begin{figure}
  \centering
  \includegraphics[]{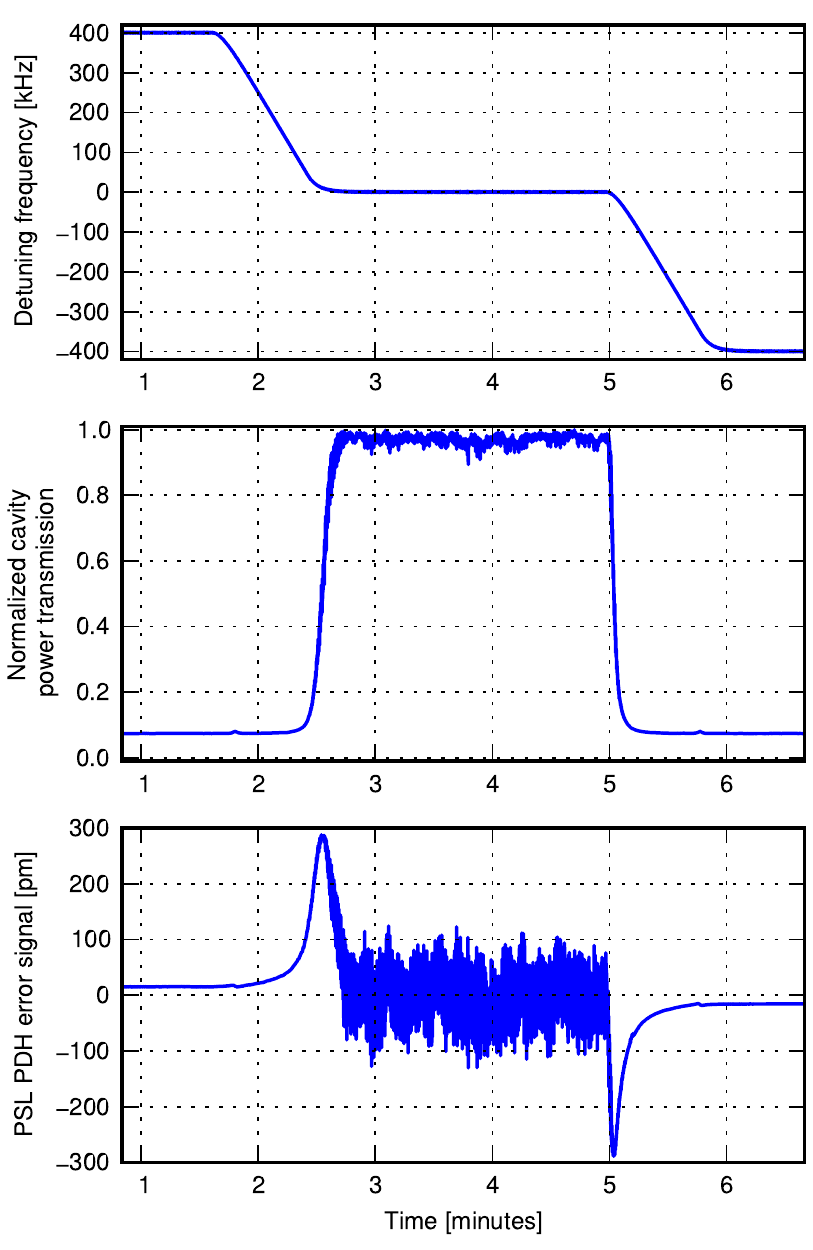}
  \caption{Sweep of cavity length control offset through the 1064~nm
    (PSL) resonance of the cavity. A detuning of 100~kHz corresponds
    to a cavity displacement of 6.7~nm.}
  \label{fig:ts}
\end{figure}

The primary usefulness of the multi-color readout is that it enables
us to precisely adjust the arm length and hold it at a desired value
independent of the state of the rest of the interferometer (as
discussed in section~\ref{sec:setup}).  Figure~\ref{fig:ts} shows a
sweep of the cavity length feedback offset through the cavity
resonance of the 1064~nm PSL beam.  The figure demonstrates that the
length detuning can be cleanly and smoothly brought to zero, at which
point the 1064~nm beam is fully resonant in the cavity.  The top plot
is the amount of the detuning in terms of the beat frequency observed
at the fine DFD output. The middle plot shows the 1064~nm intracavity
power as it passes through resonance with the cavity.  The bottom plot
shows the PSL PDH error signal.

\begin{figure*}
  \centering
  \includegraphics[width=2\columnwidth]{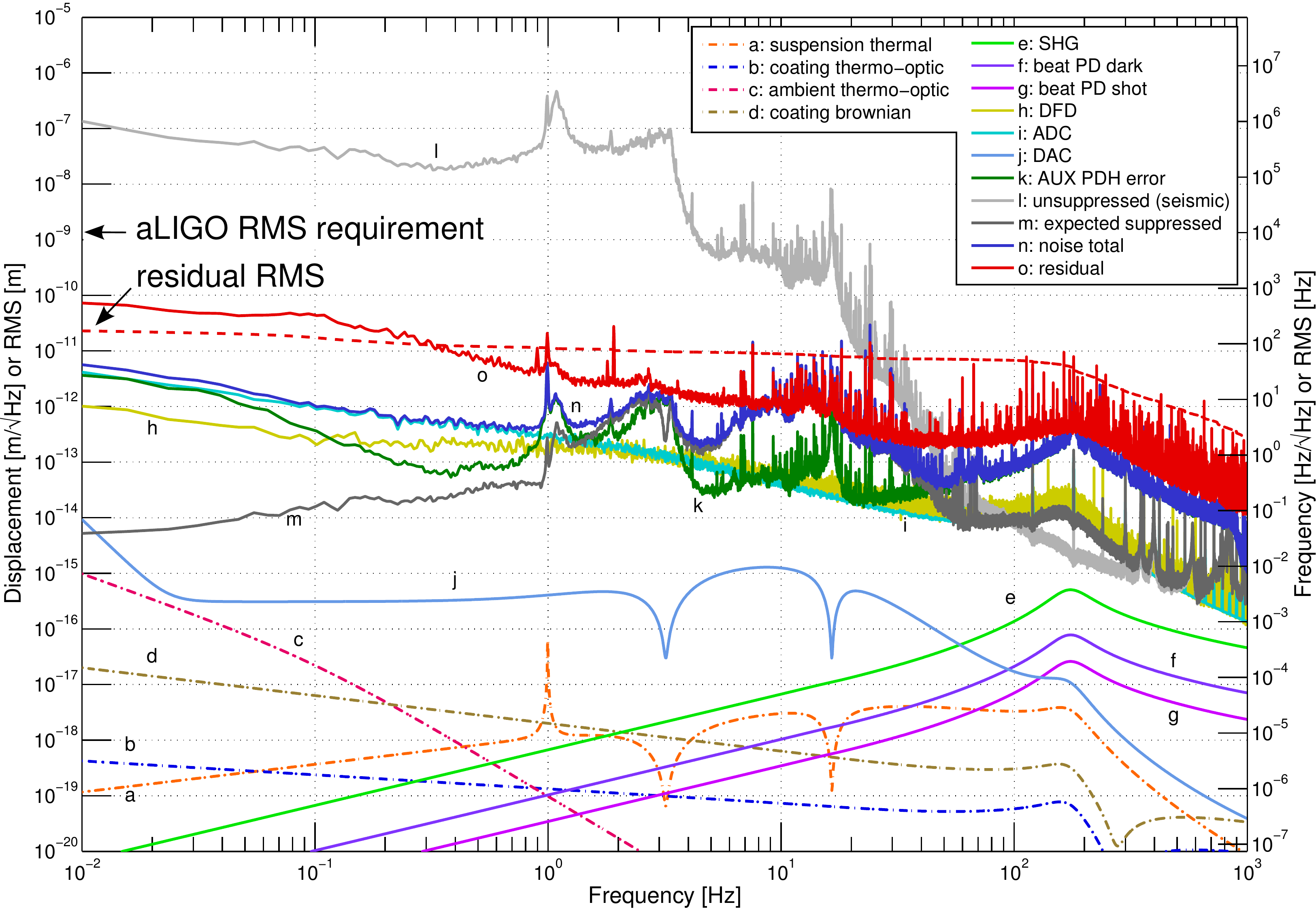}
  \caption{Residual displacement noise at 1064~nm, and noise budget of
    the locked cavity.  The right vertical axis indicates the
    corresponding frequency noise at 1064~nm.  The solid red curve is
    the overall residual cavity displacement measured by the
    out-of-loop PSL PDH detector, while the red dashed curve is the
    residual displacement RMS integrated from high frequency.  The
    colored dash-dot curves represent the estimated noise
    contributions from various fundamental noise sources, while the
    solid colored curves are the measured or estimated levels of the
    various technical noise sources.}
  \label{fig:NB}
\end{figure*}

The residual arm displacement measured in the out-of-loop PSL PDH
error signal has a root mean square (RMS) of 23.5~pm, integrated from
1~kHz to 10~mHz.  The amplitude spectral density and RMS of this
residual displacement are shown as the solid and dashed red curves
respectively in Figure~\ref{fig:NB}.  The measured RMS falls below the
aLIGO requirement of 1.3~nm RMS~\cite{T0900095}, which is based on the
the line width of the arm cavity for the 1064~nm wavelength.

Figure~\ref{fig:NB} also shows the overall noise budget of the
experiment, i.e. an accounting of all noise sources that are thought
to affect the performance of the experiment.  To determine the
contribution from a particular source we first calculate, estimate, or
measure the power spectrum of the noise, $S(f)$, at its source
(designated by a $\oplus$ in Figure~\ref{fig:model}), and propagate
the amplitude spectrum through the control model to produce an
amplitude spectrum in the ``residual displacement'' output.  The
result is
\begin{equation}
  n(f) = X(f) \sqrt{S(f)},
  \label{eq:nc}
\end{equation}
where $X(f)$ is the transfer function from the noise source to the
out-of-loop-measured residual displacement.  In the rest of this
section we will describe the contribution from each of the sources
shown in Figure~\ref{fig:NB}.

The total noise spectrum accountable from the budget is shown as the
solid blue trace in Figure~\ref{fig:NB}.  The fact that the blue trace
lies below the red trace over much of the band indicates a discrepancy
in the noise accounting.  The measured displacement spectrum is
limited by a $1/f$-shaped noise at low frequencies, and a white noise
above 100~Hz.  It is unknown at this time where these limiting factors
originate.

Table~\ref{tab:vals} in Appendix~\ref{sec:vals} describes all of the
fundamental constants, experimental values, and material properties
used in this section.

\subsection{Fundamental cavity noise sources}
This section describes various fundamental noise sources in the cavity
being measured.  While most of these noises can't be measured
directly, their levels can be estimated based on analytical models of
the underlying physics.

\subsubsection{Seismic noise}
Seismic noise, while dominant across much of the band of interest, is
suppressed by the cavity length control loop.  The light gray
``unsuppressed'' spectrum in Figure~\ref{fig:NB} is an estimate of the
free-swinging cavity motion, $\sqrt{S_{\mathrm{seis}}}$, based on the
in-loop error signal when the cavity is locked to the PSL.  This
spectrum is expected to be entirely dominated by seismic noise at
frequencies below 100~Hz, and reaches a level of roughly
$10^{-7}~\mathrm{m}/\sqrt{\mathrm{Hz}}$ below 1~Hz. The peak at 1~Hz
is due to the pendulum resonance of the optic's suspension system,
while the peak at 3~Hz is due to the resonance of the vibration
isolation stack which supports the optical table in the vacuum
chamber.

The dark gray ``expected suppressed'' trace is the ``unsuppressed''
convolved with the closed-loop transfer function from the common
cavity displacement input to the residual displacement.  This
represents the expected contribution of seismic noise to the residual
displacement once the arm length servo loop is closed.  Since this
motion is common to both laser frequencies its contribution is
suppressed by the servo to an expected level of $\sim
10^{-12}~\mathrm{m}/\sqrt{\mathrm{Hz}}$ across most of the band.
However, it is nonetheless found to be one of the main contributors to
the resultant arm displacement around 10~Hz.

\subsubsection{Suspension thermal noise}
\label{sec:susptherm}
The cylindrical cavity mirrors are suspended from a single wire loop
clamped at the top of a suspension cage.  The length of the pendulum
is 25~cm, which leads to a fundamental pendulum frequency of $f_p =
1.0$~Hz.  Suspension thermal noise originates from thermal fluctuation
of these suspension wires.  The noise is well
modeled~\cite{gonzalez2000} and its power spectrum is expressed as
\begin{equation}
  S_{\mathrm{sus}}(f) =
  \frac{4 k_B T}{(2 \pi f)^2} \mathbb{R}\left[ Y(f) \right],
\end{equation}
where $k_B$ is the Boltzmann constant, $T$ is the mean temperature,
and $f$ is the frequency.  $Y$ is the admittance of the suspended
mirror due to an external force and is described by the transfer
function
\begin{equation}
  Y(f) =
  \frac{1}{M}\ 
  \frac{\mathrm{i} f/(2 \pi f_p^2)}{1+ \mathrm{i} \phi_p - (f/f_p)^2},
\end{equation}
where $M$ is the pendulum mass, $f_p$ is the pendulum frequency, and
$\phi_p$ is the pendulum loss angle.  As with seismic noise, this
noise is common to both AUX and PSL beams so its contribution is
suppressed by the arm length servo.

The expected suspension thermal noise level in our experiment is shown
as the dot-dashed orange curve in Figure~\ref{fig:NB}.  The peak at
1~Hz is excess noise due to the pendulum resonance.  The dip at 3.2~Hz
is due to a resonant gain in the arm length servo used to suppress the
contribution from the primary mode of the vibration isolation stack,
while the dip at 16.5~Hz is resonant gain used to suppress the
contribution from the bounce mode of the optic suspension.

\subsubsection{Coating thermal noise}
Noises associated with thermal fluctuations in the mirror's
high-reflectivity coatings are an important limiting noise source in
LIGO.  While they are not expected to be a notable contribution to our
result, we touch on them here for completeness.

There are two important coating thermal noise sources:
\textit{Brownian} noise comes from thermal vibrations associated with
mechanical losses in the mirror coating.  The combined
\textit{Thermo-refractive} and \textit{thermo-elastic} noises, jointly
referred to as \textit{thermo-optic} (TO) noise, affects the laser
field as it interacts with the high-reflective coating.

Unfortunately, calculating the effect of these noises in the presence
multiple light wavelengths is not trivial.  As mentioned in
section~\ref{sec:pslpdh}, only the noises sensed differentially
between the AUX and PSL beams, and therefore not suppressed by the arm
length servo, will show up as residual displacement noise in our
experiment.  Calculating the differential effect accurately would
therefore require a fully coherent analysis at both wavelengths, which
we will not attempt here.  Instead we make the very naive assumption
that the difference between what is sensed by the AUX and PSL beams is
entirely attributable to the difference in their spot sizes on the
mirror surfaces.  We then calculate the differential thermal noise
contributions based on this differential spot area.

The Brownian thermal noise spectrum is given by~\cite{Harry:CQG2007}:
\begin{equation}
  S_{\mathrm{BR}}(f) =
  \frac{4 k_B T }{2 \pi f}
  \frac{\phi_{\mathrm{eff}} \left(1 -P^2\right)} {E \sqrt{a}},
  \label{eq:brown}
\end{equation}
where $P$ is the Poisson ratio of the substrate, $\phi_{\mathrm{eff}}$
is the effective loss angle of the coating, $E$ is the Young's modulus
of the substrate, and $a$ is the area probed.  The resultant residual
displacement spectrum from coating Browning noise in our experiment is
shown as the dot-dashed brown curve in Figure~\ref{fig:NB}.

For the thermo-optic (TO) noise contribution, we follow the coherent
treatment proposed in \cite{Matt:TOnoise}.  A Gaussian beam
illuminating a mirror senses thermal fluctuations in the coating
resulting in the noise power spectrum:
\begin{equation}
  S^{\Delta T}_{\mathrm{TO}}(f) = 
  2\sqrt{2} \frac{k_B T^2}{a \sqrt{2 \pi f \kappa s}},
  \label{eq:TOts}
\end{equation}
where $\kappa$ is the thermal conductivity, and $s$ is the heat
capacity per volume.  The overall thermo-optic noise spectrum is then
\begin{equation}
  S_{\mathrm{TO}}(f) = 
  S^{\Delta T}_{\mathrm{TO}}(f)\ 
  \Gamma_{\mathrm{tc}}
  \left(
    \chi_{\mathrm{fsm}} \Delta \bar\alpha d - \bar\beta \lambda
  \right)^2,
  \label{eq:TO}
\end{equation}
where $\Gamma_{\mathrm{tc}}$ is a correction due to the finite
thickness of the coating layers, $\Delta \bar\alpha$ is the difference
in effective thermal expansion coefficient between the coating and
substrate, $\chi_{\mathrm{fsm}}$ is a correction due to the finite
mirror size, $d$ is the thickness of the layers, $\bar\beta$ is the
effective thermo refractive coefficient, and $\lambda$ is the beam
wavelength.  The residual displacement from thermo-optic noise is
shown as the lower dot-dashed blue trace in Figure~\ref{fig:NB}.

\subsubsection{Couplings with ambient temperature fluctuations}

Potentially more significant than the inherent thermo optic noise
contribution at low frequencies is the thermo optic contribution from
low frequency ambient temperature fluctuations coupling directly to
the mirror coating.  Thermal fluctuations in the mirror coating due to
ambient temperature fluctuations in the lab can be significantly
higher than those from thermo optic excitations.  These fluctuations
dominate the thermo optic noise spectrum at low frequencies.

To estimate the thermo optic noise contribution from ambient
temperature we start with the same thermo optic noise description in
equation~\ref{eq:TO}.  But instead of using the thermo optic
fluctuations from equation~\ref{eq:TOts}, $S_{\mathrm{TO}}^{\Delta
  T}$, we instead use an estimated thermal spectrum given by:
\begin{equation}
  S_{\delta T}^{\Delta T}(f) =
  \left[
  \delta T(f)\ 
  C(f)\ 
  j(f)
  \right]^2
  ,
\end{equation}
where $\delta T(f)$ is the amplitude spectrum of the ambient
temperature fluctuations in the lab environment,
\begin{equation}
  \delta T(f) =
  3 \times 10^{-3}
  \left(\frac{0.01~\mathrm{Hz}}{f}\right)
  \frac{\mathrm{K}}{\sqrt{\mathrm{Hz}}}
  ,
\end{equation}
$C(f)$ is the transfer function through the vacuum envelope, described
by a single 0f.1~Hz pole, and $j(f)$ is the radiative transfer to the
optic surface,
\begin{equation}
  j(f) =
  \frac{4 \epsilon \sigma T^3}{2 \pi \sqrt{f \kappa \rho c}}
  ,
\end{equation}
where $\epsilon$ is the emissivity of the coating, $\sigma$ is the
Stefan-Boltzmann constant, $T$ is the mean temperature, $\kappa$ is
the thermal conductivity of the substrate, $\rho$ is the density of
the substrate, and $c$ is the specific heat capacity.  The
contribution from this affect is shown as the pink dot-dashed curve in
Figure~\ref{fig:NB}.

\subsection{Technical noise sources}

This section describes the contribution from various technical noise
sources that can be measured directly in the experiment.

\subsubsection{Laser frequency noise}
Frequency noise associated with the AUX and PSL lasers is generally
suppressed by the control loops that keep the lasers locked to the
main arm and mode cleaner cavities.  However, since all control loops
are coupled together at some level, there is a possibility of laser
frequency noise contributing to the measured residual displacement
noise.

As discussed in section~\ref{sec:pslmc} frequency noise from the PSL
is significantly suppressed and can therefore be ignored.  However,
coupling from the AUX laser is at a much higher level.  We estimate
its contribution by observing the residual noise in the AUX PDH error
signal while the AUX-cavity loop is locked.  We then assume that this
noise is due almost entirely to unsuppressed laser frequency noise
fluctuations.  The resultant contribution from this noise is the dark
green ``AUX PDH error'' trace in Figure~\ref{fig:NB}.

\subsubsection{Second-harmonic generation noise}
Noise due to the second-harmonic generation process is assumed to be
added to the frequency noise of the frequency-doubled laser beam.  The
upper limit of the noise level is assumed to be $1\times 10^{-5} f\
\mathrm{Hz}/\sqrt{\mathrm{Hz}}$~\cite{dmass:ppktp}.  
Laser frequency doubling happens in two places in our experiment: in
the AUX laser output and on the PSL beam for the beat note detection.
The contribution from the PSL doubling in the beat note detection is
much more significant, so it is this level that is shown as the light
green curve in Figure~\ref{fig:NB}.

\subsubsection{Shot noise and detector dark noise}
Both shot noise and dark noise appear as white noise (in the detection
band) in the broadband RF photodetectors used in the experiment.  When
measuring the frequency of a signal, the measured voltage noise can be
converted to frequency noise on the detected signal
by~\cite{Wolaver1991}:
\begin{equation}
  S_{\mathrm{PD}}(f) = \frac{2 S_{\mathrm{V}}(f)}{V_{\mathrm{RF}}^2} f^2,
\end{equation}
where $S_{\mathrm{V}}$ is the input-referred voltage noise, and
$V_{\mathrm{RF}}$ is the voltage amplitude of the main RF signal.

The most dominant contribution from these noises comes from the beat
note detection photodetector.  The dark current noise level of the
photodetector used is $12~\mathrm{pA}/\sqrt{\mathrm{Hz}}$ between
10~MHz and 80~MHz.  The incident power on the PD is
$200~\mu\mathrm{W}$ which produces $60~\mu\mathrm{A}$ of DC
photocurrent, corresponding to a shot noise level of
$4~\mathrm{pA}/\sqrt{\mathrm{Hz}}$.  The resultant frequency noise
spectra at the detector input for these noise sources are shown as the
purple (dark) and magenta (shot) traces in Figure~\ref{fig:NB}.

\subsubsection{Frequency discriminator noise}
The comparator in the delay-line frequency discriminator adds white
noise during the process of reshaping RF signals into square waves.
The noise level is measured from the output of the DFD while being
driven by a pure RF sine wave.  The level, referred to the input of
the DFD, was found to have a total contribution of $10^{-14} -
10^{-13}\ \mathrm{m}/\sqrt{\mathrm{Hz}}$ after applying the loop
correction factor (olive curve in Figure~\ref{fig:NB}).

\subsubsection{ADC noise}
ADC noise is easily measured directly by terminating the inputs to the
analog filters that whiten the signal before digitization, and then
measuring the spectrum digitally.  The affect of the of the ADC
whitening is compensated for within the digital system.  The resultant
contribution, referred to the input of the DFD, sees the same loop
correction factor as the DFD (cyan trace in Figure~\ref{fig:NB}).

\subsubsection{DAC noise}
DAC noise is directly measured by digitally generating a 3~Hz signal,
representing the peak frequency of the error signal while locked, and
then measuring the output noise spectrum.  The resultant noise
contribution is mostly flat at a level of about $1~ \mu \mathrm{V}/
\sqrt{\mathrm{Hz}}$ and is shown as the pale blue curve in
Figure~\ref{fig:NB}).  The dips at 3.2~Hz and 16.5~Hz are due to the
affect of the resonant gains stages discussed in
section~\ref{sec:susptherm}.

\subsection{Scaling noise sources for Advanced LIGO}
In this section we look briefly at how various noise sources in our
experiment can be scaled to Advanced LIGO.  We find that certain noise
contributions will be more prominent in aLIGO, but that they are
addressed in the aLIGO design such that they should not pose a
significant problem.

\subsubsection{Frequency noise}
From the relation between frequency noise and length fluctuations
expressed in equation~\ref{eq:f2l} we can see that the 100-times
longer arm cavities of aLIGO means that aLIGO will be 100 times more
sensitive to laser frequency fluctuations.  For a displacement noise
requirement of 1~nm RMS, the beat note frequency stability requirement
goes from 8.8~kHz in the 40m experiment to 83~Hz in aLIGO.  This puts
a much stricter requirement on the frequency noise of the AUX laser
and SHG noises.

Advanced LIGO will mitigate this issue in a couple of different ways.
First, aLIGO will phase lock the AUX laser to the PSL frequency
through the use of a fiber-based phase-locked loop (PLL).  This will
improve noise below 50~Hz, while making it worse at high frequencies.
The increased high frequency noise can then be addressed through
optimization of the servo controls.  The gain of the AUX-cavity loop
can be increased to suppress the excess noise from the AUX laser, and
the bandwidth of the arm length stabilization loop can be decreased so
that any residual noise will not be injected into the cavity motion.

\subsubsection{Readout and frequency discriminator noise}
Readout and electronics noise sources should become less severe in
aLIGO.  The interferometer response will generally grow in proportion
to length, resulting in a higher SNR against these noise sources.

The frequency discriminator, on the other hand, generally does not
scale with the base line length since it reads out the frequency of
the beat note rather than the optical phase.  For this reason the
readout noise of the discriminator will make a 100 times larger
contribution to the noise budget than it does in our experiment.  This
could likely be the limiting noise source with a frequency noise level
of 1~Hz/$\sqrt{\mathrm{Hz}}$ at 10 Hz.  The situation can be improved
by using a small-range discriminator such as a VCO-based PLL or a
longer cable in the DFD.

\subsubsection{Seismic length fluctuations}
Length fluctuations due to seismic noise should become somewhat easier
to handle in aLIGO since the test masses will be far more isolated
from ground vibration due to sophisticated aLIGO seismic isolation
systems.  Depending on how large the residual seismic fluctuations are
the unity gain frequency of the arm length servo loop should be able
to be lowered.  This is generally good since it avoids injection of
undesired control noises at high frequencies.

\section{Future Work}
\label{sec:future}

Besides precise control of suspended Fabry-Perot cavities,
multi-wavelength readout also has the potential to improve performance
of other optical systems.
Here we present futuristic ideas which can potentially reduce the
fundamental noise sources such as quantum noise and mirror thermal
noise through the use of multi-wavelength readout, as well as an idea
to precisely characterize an optical cavity.

\subsection{Multi-wavelength readout for manupulating the quantum
  noise limit}

By resonating multiple laser beams with different wavelengths in a
single interferometer, traditional quantum noise limits can
potentially be modified.

One example is cancellation of quantum back-action for GW detectors
~\cite{PhysRevD.76.062002}.  Imagine a main carrier field resonating
with high power in an interferometer arm cavity, and a low-power
auxiliary laser beam with a different wavelength resonant only in the
interferometer vertex (anti-resonant in the arm cavities).  The
high-power main carrier field would produce quantum radiation pressure
noise on the test masses.  The low-power auxiliary laser beam, on the
other hand, would sense only the differential motion of the two input
test masses, and therefore not be sensitive to gravitational wave
signals.  An optimal combination of the two carriers beam with Wiener
filters could then be used to cancel the low-frequency back action
noise, while not losing information from gravitational waves.

Another idea is to resonate both carrier wavelengths in the arm
cavities. The design of the optics could be made such that the optical
properties for the different wavelengths are different leading to
different frequency sensitivities for the two beams.  For example, the
input test masses could have higher transmittance for one wavelength
over the other.  By tuning different wavelengths and optimally
combining their outputs, one may be able to shape the quantum noise
spectrum in a much more flexible way than in the single wavelength
case.

Multiple beams with different wavelengths could also potentially be
used to manipulate the dynamics of test masses in optical cavities. In
particular, it is well known that the multi-bounce laser fields in
optical cavities modify the dynamics of the cavity mirrors via
radiation pressure~\cite{BuCh2002}.  Under appropriate conditions this
can result in modifications of the opto-mechanical coupling and a
higher response against optical phase changes.  For example, in
gravitational wave detectors that use signal-recycling cavities, two
wavelengths of light appropriately detuned from the resonance of the
signal-recycling cavity can result in a radiation pressure force that
reduces the effective inertia of the test mass at low frequencies.
This can significantly amplify the response of the interferometer to
gravitational-wave signals~\cite{PhysRevD.83.062003}.  It is therefore
possible that a multi-wavelength technique could allow for surpassing
the standard quantum limit over a broad frequency band.

\subsection{Thermal noise estimation}

In a frequency regime where sensitivity is strongly limited by mirror
thermal noise, it may be possible to use two different laser
wavelengths to differentially sense the thermal noise in the two
fields.  It may then be possible to combine signals from the two
lasers in order to yield one data stream representing the cavity
length fluctuations and another with purely the thermal noise. This
technique may be capable of giving a moderate decrease in the
effective thermal noise.  However, the differential frequency noise
between the wavelengths would need to be investigated more precisely.

\subsection{Precise cavity mode characterization}

Since multi-wavelength metrology enables us to detune the laser
frequency from one of the cavity resonances in a quasi-static manner,
various longitudinal and spacial characteristics of the cavity can be
precisely inspected.  Precise scanning of a resonance can provide a
measure of cavity finesse, while scanning over multiple free spectral
ranges gives us a measurement of the absolute length of the cavity.
The frequency spacing of transverse spatial modes can be obtained by
inspecting the resonances of the fundamental and higher-order modes,
therefore providing information about the cavity geometry and the
figure error of the cavity optics.  The power transmitted during
scanning can also tell us the mode matching efficiency between an
incident beam and cavity eigenmodes.

\section{Conclusions}
\label{sec:conclusion}
Using multiple lasers, we have demonstrated a tractable strategy for
sensing the cavity lengths of a complex interferometer for
gravitational wave detection. This method can be replicated and
applied to any of the detectors in the upcoming worldwide network
(\mbox{LIGO}, \mbox{Virgo}, \mbox{KAGRA}).

The noise limits are now well understood and well below the
requirements necessary for aLIGO.  The mirror masses for these new
detectors are 30-40 kg and they have thermal time constants of several
hours. Long periods spent without the interferometer locked introduce
enormous thermal transients in the interferometer which perturb the
delicate operating state.  The use of this technique should allow for
a significantly higher duty cycle in the future.

In addition, the technique has promise to improve the sensitivity of
the next generation of interferometers through the use of
multi-wavelength readout to partially cancel some of the thermal and
quantum noises which limit the more conventional designs.

\appendix
\section{Control Model Transfer Functions}
\label{sec:zpk}

\begin{figure}
  \centering
  \includegraphics[width=\columnwidth]{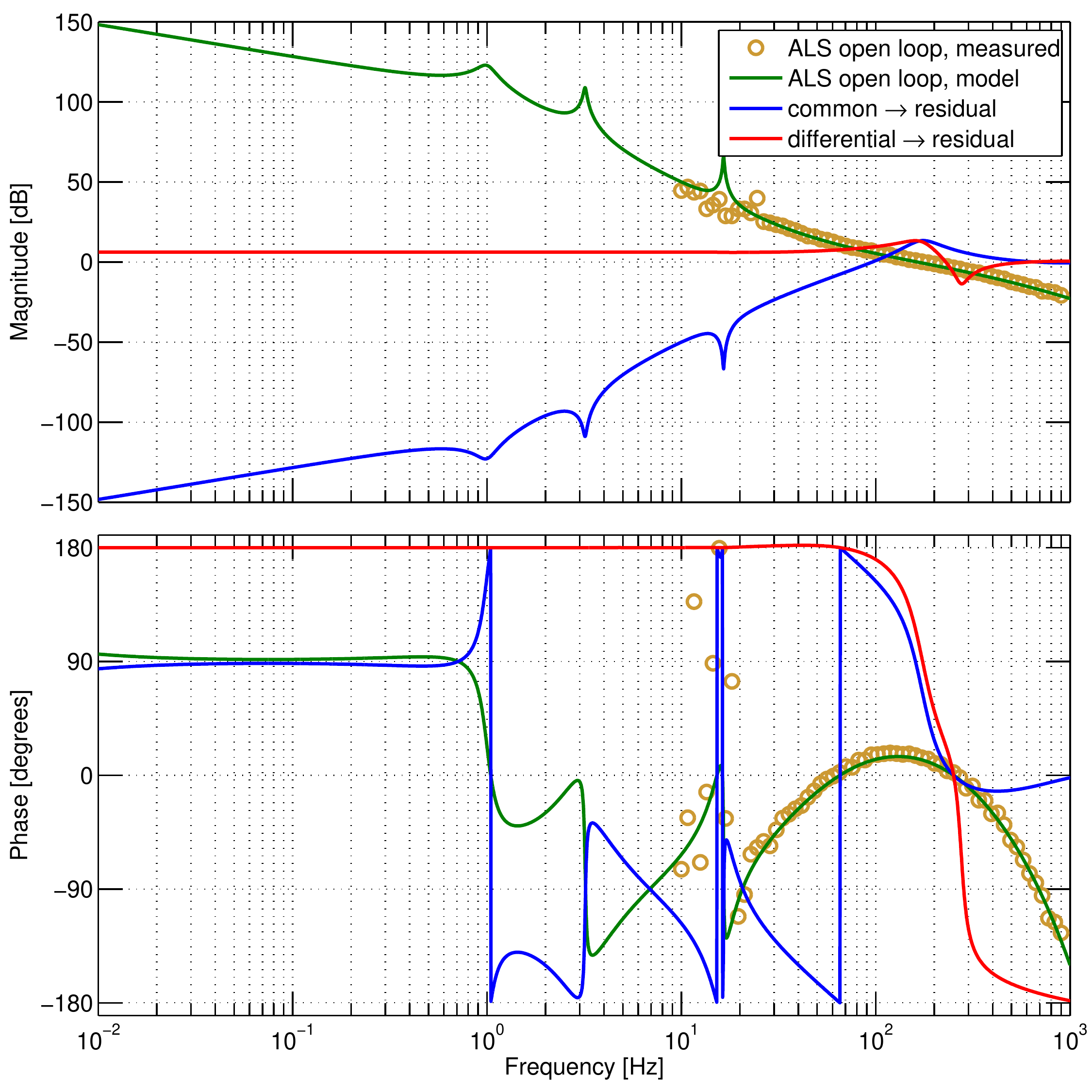}
  \caption{Bode plot of control model transfer functions.  The green
    trace is the full open-loop transfer function of the arm length
    servo control loop.  The blue and red traces are the transfer
    functions from external disturbances to residual displacement, for
    AUX/PSL common and differential sensing.}
  \label{fig:bode}
\end{figure}

Table~\ref{tab:zpk} lists all of the zeros, poles, and gains for the
various control elements in the control model described in
section~\ref{sec:model} and Figure~\ref{fig:model}.  For a system with
$p_m$ poles and $z_n$ zeros (both specified in Hz) and gain $k$, the
transfer function would be given by:
\begin{equation}
  X(f) =
  k\ 
  \frac{\prod_n (1 + \mathrm{i}f/z_n)}{\prod_m (1 + \mathrm{i}f/p_m)}
  ,
\end{equation}
Figure~\ref{fig:bode} shows Bode plots (amplitude and phase) of a
couple of the key transfer functions from the full control model.

In addition to the blocks represented in Table~\ref{tab:zpk}, the
blocks labeled ``$L\rightarrow\nu$'' in Figure~\ref{fig:model}
represent the conversion from displacement ($dL$) to frequency
($d\nu$) described in equation~\ref{eq:f2l}, i.e.:
\begin{equation}
  L\rightarrow\nu = \frac{\nu}{L},
\end{equation}
where $\nu = c/\lambda$.

\begin{table}[t]
\begin{center}
\begin{tabular}{c|c|c|c}
  \hline
  \hline
  \bf element & \bf zeros (Hz) & \bf poles (Hz) & \bf gain \\
  \hline
  $C_{\mathrm{AUX}}$ & - & 18.5k & 1 \\
  $D_{\mathrm{AUX}}$ & - & - & 5.0e-6 \\
  $F_{\mathrm{AUX}}$ & 1.0, 100, 10k & 0.1m, 1.2, 2.0 & 2.1e8 \\
  $A_{\mathrm{AUX}}$ & - & 100k & 5.0e6 \\

  $D_{\mathrm{DFD}}$ & - & - & 2.16e-7 \\
                     & 40, 40, & 1.0m, 500, & \\
  $F_{\mathrm{ALS}}$ & $1.655 \pm 2.739\mathrm{i}$, & $0.052 \pm 3.200\mathrm{i}$, & 1.0e6 \\
                     & $3.320 \pm 16.163\mathrm{i}$ & $0.105 \pm 16.500\mathrm{i}$ & \\

  $A_{\mathrm{TM}}$  & - & $0.1 \pm 0.995\mathrm{i}$ & 8.0e7 \\

  $C_{\mathrm{PSL}}$ & - & 3.8k & 1 \\
  $D_{\mathrm{PSL}}$ & - & - & 2.5e-5 \\
  $F_{\mathrm{PSL}}$ & 4k, 20k, 20k & 40, 1k, 1k & 2.3e4 \\
  $A_{\mathrm{PSL}}$ & - & 100k & 5.0e6 \\

  SHG                & - & - & 2 \\
  \hline
  \hline
\end{tabular}
\caption{Zeros, poles and gain of the control model blocks.}
\label{tab:zpk}
\end{center}
\end{table}

\section{Symbol Definitions and Values}
\label{sec:vals}

Table~\ref{tab:vals} shows the values used for various variables in
the text, including all fundamental constants, experimental values,
and material properties.

\begin{table}[t]
\begin{center}
\begin{tabular*}{\columnwidth}{c|p{4cm}|c|c}
  \hline
  \hline
  \bf symbol & \bf name & \bf value & \bf SI unit \\
  \hline

  $k_B$ & Boltzmann's constant & 1.38e-23 & J\ K$^{-1}$ \\
  $\sigma$ & Stefan-Boltzmann constant & 5.67e-8 & W\ m$^{-2}$K$^{-4}$ \\

  $T$ & mean temperature & 290 & K \\
  $\lambda$ & beam wavelength & 1.064e-6 & m \\

  $M$ & cavity mirror pendulum mass & 0.243 & kg \\
  $f_p$ & pendulum frequency & 1.0 & Hz \\
  $\phi_p$ & pendulum loss angle & 1.7e-4 & -- \\

  $P$ & Poisson ratio of (substrate) & 0.167 & -- \\
  $E$ & Young's modulus (substrate) & 7.27e10 & N\ m$^{-1}$ \\

  $\kappa$ & thermal conductivity & 1.38 & W\ m$^{-1}$\ K$^{-1}$ \\
  $s$ & heat capacity per volume & 1.62e6 & m$^{-3}$\ K$^{-1}$ \\
  $c$ & specific heat capacity & 740 & J\ kg$^{-1}$\ K$^{-1}$ \\
  $\epsilon$ & emissivity & 0.9 & -- \\
  $\rho$ & density & 2202 & kg\ m$^{-3}$ \\

  \hline
  \multicolumn{4}{c}{ITM} \\
  \hline
  $\phi_{\mathrm{eff}}$ & effective coating loss angle & 8.65e-8 & -- \\
  $\Delta \bar\alpha$ & effective thermal expansion difference & 3.59e-6 & K$^{-1}$ \\
  $\bar\beta$ & effective thermal refraction & 2.35e-6 & K$^{-1}$ \\
  \hline
  \multicolumn{4}{c}{ETM} \\
  \hline
  $\phi_{\mathrm{eff}}$ & effective coating loss angle & 1.24e-7 & -- \\
  $\Delta \bar\alpha$ & effective thermal expansion difference & 4.54e-6 & K$^{-1}$ \\
  $\bar\beta$ & effective thermal refraction & 1.11e-6 & K$^{-1}$ \\

  \hline
  \hline
\end{tabular*}
\caption{Values of fundamental constants, and material properties and
  variables for the 40m mirror coatings.  If not specified, material
  properties are for the mirror coating.}
\label{tab:vals}
\end{center}
\end{table}

\section*{Acknowledgments}
We gratefully acknowledge illuminating discussions with Bram
Slagmolen, Nicolas Smith-Lefebvre and Peter Fritschel.  We also thank
the National Science Foundation for support under grant PHY-0555406.


\end{document}